\documentclass[page-classic]{epl2} 

\usepackage{graphicx}
\usepackage{amsmath}
\usepackage{amssymb}

\def\110{$\langle 110 \rangle$}
\def\d111{$\langle 111 \rangle$}
\def\211{$\langle 211 \rangle$}
\def\s2g{$S^2G$}
\def\m{$_{\mathrm{max}}$}
\def\ztm{$zT_{\mathrm{max}}$}
\def\T{\mathcal{T}}

\title{Prediction of high $zT$ in thermoelectric silicon nanowires with axial germanium heterostructures}
\shorttitle{High $zT$ thermoelectric Si nanowires with axial Ge heterostructures} 

\author{M. Shelley \and A. A. Mostofi\thanks{Email: \email{a.mostofi@imperial.ac.uk}}}
\shortauthor{Matthew Shelley and Arash A.~Mostofi}

\institute{                    
  The Thomas Young Centre for Theory and Simulation of
  Materials, Imperial College London, London SW7 2AZ, UK\\
}


\abstract{We calculate the thermoelectric figure of merit, $zT=S^2G
T/(\kappa_l+\kappa_e)$, for $p$-type Si nanowires with axial Ge
heterostructures using a combination of first-principles
density-functional theory, interatomic potentials, and
Landauer-Buttiker transport theory. 
We consider nanowires with up to
8400 atoms and twelve Ge axial heterostructures along their
length. We find that introducing heterostructures always
reduces \s2g, and that our calculated increases in $zT$ are
predominantly driven by associated decreases in $\kappa_l$. 
Of the systems considered, \d111 nanowires with a regular distribution
of Ge heterostructures have the highest figure-of-merit:
$zT=3$, an order of magnitude larger than the equivalent pristine
nanowire. 
Even in the presence of realistic structural disorder, in the
form of small variations in length of the heterostructures,
$zT$ remains several times larger than that of the
pristine case, suggesting that axial heterostructuring is a
promising route to high-$zT$ thermoelectric nanowires. 
}

\begin{document}

\maketitle

In recent years, silicon nanowires (SiNWs) have been proposed for use as
chemical sensors~\cite{cui_2001}, photovoltaics~\cite{garnett_2010} and
thermoelectrics~\cite{hochbaum_2008, boukai_2008}.
Hicks and Dresselhaus~\cite{hicks_1993} first identified that NWs could 
be used to improve the thermoelectric figure of merit, $zT$, over bulk
and two-dimensional superlattice (2DSL) values. While the dramatic
increases that were predicted have not been yet realised, much progress
has been made: Refs.~\cite{hochbaum_2008,boukai_2008} report $zT\sim1$
for SiNWs, a 100-fold increase over the value for bulk Si.

The measure of the performance of a thermoelectric material is given by its
figure of merit $zT=S^2G T/(\kappa_l+\kappa_e)$, 
where $S$, $G$ and $T$ are the Seebeck co-efficient, electronic 
conductance and average temperature of the two contacts, respectively, and
$\kappa_l$ and $\kappa_e$ are the lattice and electronic contributions to the thermal 
conductance, respectively.
$zT$ may, therefore, be increased by designing materials that have
either higher thermoelectric power factor \s2g, or lower thermal
conductivity. The relatively high $zT$ seen in recent experiments on
SiNWs has been attributed to both of these effects: an increase in $S$
resulting from enhanced phonon drag~\cite{boukai_2008}, and a decrease
in $\kappa_l$ resulting from surface scattering of
phonons~\cite{hochbaum_2008, li_sinw_2003}. 

By analogy with 2DSLs, which show a reduction in $\kappa_l$ compared
to its value both in the bulk and in the alloy
limit~\cite{borca_2000}, superlatticed NWs have been
proposed as a possible route toward high $zT$
thermoelectrics~\cite{lin_2003}, through both enhancement of the power
factor \s2g\ and reduction of $\kappa_l$, as compared to pristine
NWs. 
Experimental evidence for Si-SiGe
NWs~\cite{li_slnw_2003} supports the idea that superlatticing
results in a reduction in $\kappa_l$, although direct experimental
evidence for increased \s2g\ is still missing.

\begin{figure}
\onefigure[width=8.5cm]{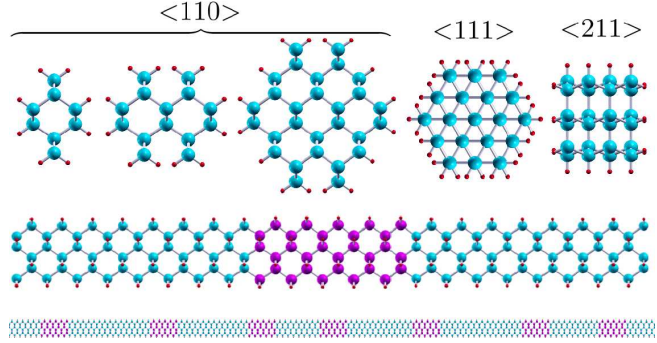}
\caption{Top: Cross-sections of SiNWs. Labels indicate the crystal
  direction of the longitudinal ($z$) axis (pointing into the page). 
Diameters (left-to-right): 0.78~nm, 1.02~nm, 1.44~nm, 1.14~nm and
1.06~nm. 
Middle: A single Ge heterostructure in a
SiNW (Ge and Si atoms in magenta and blue,
respectively). Bottom: A multiple heterostructure nanowire (MHNW) with
an arbitrary distribution of Ge heterostructures.}  
\label{fig:NW_systems}
\end{figure}

In this Letter, we use a combination of first-principles
density-functional theory simulations and calculations with
interatomic potentials in order to compute $zT$ for axially
heterostructured Si-Ge NWs
within the coherent transport regime using the
Landauer-Buttiker approach~\cite{landauer_1970,buttiker_1985}. 
More specifically, we calculate $zT$ at 300~K for thin ($<2$~nm
diameter), $p$-type \110, \d111 and \211 H-passivated SiNWs
(Fig.~\ref{fig:NW_systems}, top) containing: (a) single axial
Ge heterostructures with lengths ranging from 0.4~nm to 4.3~nm
(Fig.~\ref{fig:NW_systems}, middle); (b) 
multiple Ge heterostructures of uniform length, distributed
along the length of the NW either randomly, periodically, or as
a Fibonacci chain (Fig.~\ref{fig:NW_systems}, bottom); and (c)
multiple Ge heterostructures whose lengths approximately
follow a Gaussian distribution. 

Our calculations demonstrate that: (i) the introduction of a single Ge
heterostructure in a SiNW can lead to a 3.5-fold increase in $zT$ as
compared to the equivalent pristine SiNW; (ii) this can be further
enhanced to a 7.4-fold increase by introducing multiple Ge
heterostructures and by controlling their spatial distribution along
the length of the SiNW; (iii) this observed enhancement in $zT$ is
almost entirely due to a reduction in thermal conductivity rather than
an increase in the power factor \s2g -- indeed, we find that
introducing Ge heterostructures in an SiNW always results in a
decrease in \s2g; and (iv) introducing further disorder, in the form
of a variability in the length of the Ge heterostructures within a
SiNW, results in a decrease in $zT$ as compared to the case in which
the heterostructures are all identical, highlighting the importance of
precise atomic-scale control that may be required in order to
fabricate high-$zT$ NWs.

Our method (described in detail
elsewhere~\cite{shelley_method2011}), in which accurate yet compact
model Hamiltonians of large-scale systems are constructed from
first-principles calculations, enables us to study transport through
meso-scale systems with modest computational cost: our largest
simulations consist of a conductor region of length 116~nm (8432
atoms) coupled to semi-infinite leads. Our procedure is
largely automated, which has made it possible to perform high-throughput
calculations and undertake a comprehensive study of a large structural
parameter space. With little modification, our general approach may be
easily used to calculate transport properties in other quasi one-dimensional
systems.

We note that analogous approaches have been used recently
  for calculating electronic and/or thermal transport in large-scale
  1D systems, e.g., in SiNWs~\cite{markussen-prb06,markussen-prl09} and
  carbon nanotubes~\cite{lee-prl05,rocha-prl08,savic-prl08}.

\section{Electronic transport properties}
Starting from a plane-wave density-functional theory (PW-DFT)
calculation\footnote{We use the Quantum-Espresso
  package\cite{giannozzi_2009}, the local-density approximation for
  exchange and correlation, norm-conserving pseudopotentials, a 400~eV
energy cut-off for the PW basis set, and $\Gamma$-point sampling of
the Brillouin zone.}, a unitary transformation is
applied to the extended  ground state eigenfunctions in order to obtain
maximally-localized Wannier functions (MLWFs)~\cite{marzari_1997}
and, hence, the Hamiltonian matrix in the basis of MLWFs. 
Due to the localized nature of MLWFs in real-space, the Hamiltonian matrix can be
spatially partitioned and used in so-called ``lead-conductor-lead''
Landauer-Buttiker~\cite{landauer_1970,buttiker_1985} transport calculations,
using standard Green function techniques~\cite{caroli_1971,
  lee_1981,meir_1992,nardelli_2001,shelley_method2011}. 
In our case, the semi-infinite ``leads'' are pristine (H-passivated) SiNWs and the 
central ``conductor'' region comprises of SiNW with some axial distribution
of Ge heterostructures. An example of  such a system exhibiting a
single Ge heterostructure is shown in Fig.~\ref{fig:NW_systems}
(middle panel).
Although we begin with PW-DFT calculations with periodic boundary conditions,
we determine electronic transport properties under open boundary
conditions.
Once the electronic density of states of the 
conductor and the transmission function $\T(\epsilon)$ are calculated, 
one can write~\cite{sivan_1986,esfarjani_2006} 
$G=e^2L_0(\mu)$, $S=L_1(\mu)/eTL_0(\mu)$ and 
$\kappa_e=\frac{1}{T}\{L_2(\mu)-[L_1(\mu)]^2/L_0(\mu)\}$, where
\begin{equation}
L_m(\mu)=\frac{2}{h}\int^{\infty}_{-\infty}\T(\epsilon)(\epsilon-\mu)^m
\left(\frac{-\partial
    f(\epsilon,\mu)}{\partial\epsilon}\right)d\epsilon ,
\label{eq:L_m}
\end{equation}
and $f(\epsilon,\mu)=1/\{\exp[(\epsilon-\mu)/k_BT]+1\}$
is the Fermi-Dirac function at chemical potential $\mu$. In this Letter, we 
focus on hole transport, so that we can associate $\mu$ with a carrier concentration
that is driven by $p$-doping\footnote{This doping is `artificial' in the sense 
that we do not directly include dopant atoms in our calculation.}.

As full DFT structural relaxation of our large
  heterostructured NWs would have been computationally
  intractable, the atomic configurations used for the 
  electronic calculations were built by piecing together
  unit cells of pristine Si and Ge NWs whose equilibrium lattice
  parameters\footnote{These were 3.775~\AA\ (3.910~\AA), 9.224~\AA\ (9.497~\AA),
    and 6.470~\AA\ (6.692~\AA), for our largest diameter Si (Ge) NWs in
    the \110, \d111 and \211 growth directions, respectively.} were
  calculated separately with DFT. For the smallest  
  NWs considered, the results obtained from our approach had only
  small quantitative and no qualitative difference as compared to
  those from the equivalent fully relaxed
  structure~\cite{shelley-phd11}.

\begin{figure}
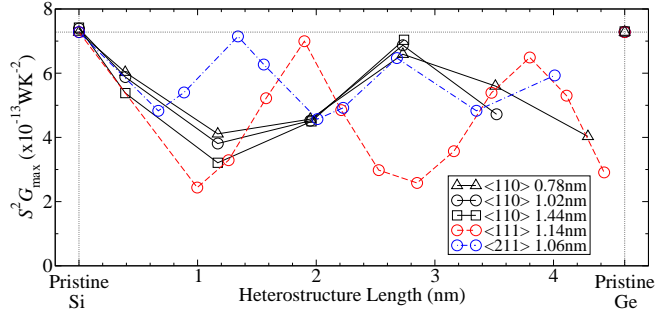

\onefigure[width=8.5cm]{sing_het_S2G.eps}
\caption{Maximum thermoelectric power factor \s2g\m\ at 300~K for single 
Ge heterostructures in \110 (black solid lines), \d111 (red dashed lines)
and \211 (blue dot-dashed lines) SiNWs. Three diameters 
0.78~nm (triangles), 1.02~nm (circles) and 1.44~nm (squares) are shown for
the \110 direction.
Pristine SiNWs are shown plotted as zero heterostructure length and
pristine Ge NWs values are shown on the right.}
\label{fig:sing_het_S2G}
\end{figure}
Fig.~\ref{fig:sing_het_S2G} shows the maximum\footnote{As can be seen
  from Eq.~(\ref{eq:L_m}), the electronic transport coefficients are
  functions of chemical potential $\mu$. Throughout this work, \s2g\m\
  and \ztm\ are defined to be the maximum values of the power factor
  and the figure of merit, respectively, as a function of $\mu$. The
  maximal power factors shown in Fig.~\ref{fig:sing_het_S2G} are
  obtained, in all cases, for values of $\mu$ within 30~meV of the Si
  valence band edge.}
thermoelectric power factor \s2g\m\ for a range of Ge heterostructure 
lengths in \110 (black solid lines), \d111 (red dashed line) and \211
(blue dot-dashed line) SiNWs.
We note first that in no case does the introduction of an axial
heterostructure result in an increase of \s2g\m, the value of which
is, at best, approximately the same as that of a pristine SiNW. 
The similarity of the results between pristine SiNWs and GeNWs  
is also interesting to note. Since all the NWs investigated only have
a single channel that is available for conduction at the top of the 
valence manifold, we confirm that the most important factor for
\s2g\m\ in quasi-one-dimensional systems is the number of conducting 
channels at this edge~\cite{kim_2009}. 

The oscillations 
  in Fig.~\ref{fig:sing_het_S2G} can be explained by a model in
  which the heterostructure is considered as a 1D quantum
  potential well of width $L$, corresponding to the length of the
  heterostructure, and depth $V_0$, corresponding to the band offset
  between Si and Ge. The reflection amplitude of a wave incident on
  such a well vanishes when the well-known Fabry-Perot resonance
  condition is satisfied, $qL=n\pi$, where $q$ is the wavevector inside
  the well and $n$ is an integer. For holes entering
  the heterostructure, this condition gives
$E_n(L)=-\frac{\hbar^2}{2m^{\ast}}\frac{n^2}{L^2} + V_0$,
where $m^{\ast}$ is the effective mass of holes and $E_n(L)$ 
are energies at which resonances occur in the
transmission. For the \d111 NWs, which show the strongest
oscillations in \s2g\m, plotting the resonance energies against
$1/L^2$ produces an excellent linear fit (not shown), with correlation
coefficient $r^2=0.998$, giving $m^{\ast}=0.28~m_{\rm e}$ and
$V_0=0.32$~eV. Models such as this may be a useful additional tool for
the optimization of heterostructure lengths in thermoelectric
devices~\cite{wang-prb09}.

\section{Phononic transport properties}
We determine the lattice thermal conductance $\kappa_l$ in an analogous
way to the electronic conductance, by
segmenting the system into a lead-conductor-lead geometry. 
Instead of finding the Green function of a Schr\"{o}dinger-type eigenvalue  
problem, we determine the Green function that solves the eigenvalue
problem relating nuclear displacements $u$ to the dynamical matrix $K$
and phonon frequency $\omega$: $Ku=\omega^2u$. Applying the thermal equivalent of the 
Caroli formula~\cite{caroli_1971, wang_epjb_2008}, we obtain the phonon transmission
function, $\T(\omega)$, in the limit of non-interacting
phonons and coherent phonon transport. 
One can then write~\cite{yamamoto_2006}
\begin{equation}
\kappa_l=\frac{\hbar^2}{2\pi k_BT^2}\int^\infty_0\T(\omega)\omega^2
\frac{e^{\hbar\omega/k_BT}}{(e^{\hbar\omega/k_BT}-1)^2}d\omega.
\label{eq:kappa_l}
\end{equation}
Determination of the dynamical matrix 
using first-principles methods is computationally intractable for
the large NW supercells with heterostructures that are
considered here. To structurally relax the NWs
and obtain their dynamical matrices, therefore, we use Tersoff  
potentials~\cite{tersoff_1988_89}, which have been shown to
give accurate values for lattice thermal conductivities for thin
pristine SiNWs, as compared to DFT
calculations~\cite{markussen_2008}. 
It is worth noting that
the approach outlined above neglects Umklapp scattering, which would
further decrease $\kappa_l$ at the temperature with which we are
concerned (300~K). 

\begin{figure}
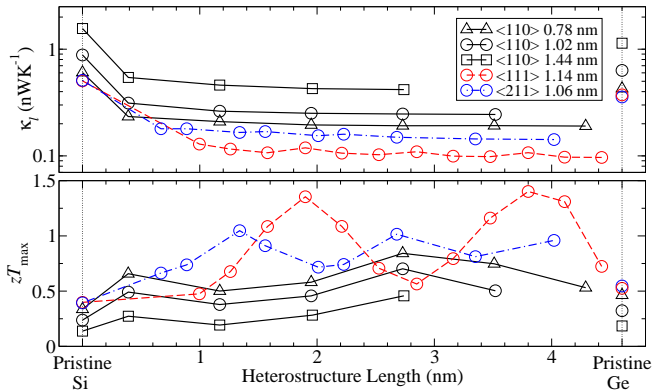

\onefigure[width=8.5cm]{sing_het_kl_zT.eps}
\caption{Dependance of $\kappa_l$ (top panel) and \ztm\ (bottom panel)
at 300~K as a function of Ge heterostructure length for \110, \d111 
and \211 SiNWs. Labelling is equivalent to  Fig.~ \ref{fig:sing_het_S2G}, 
again plotting pristine SiNWs as zero heterostructure length and pristine 
Ge NWs also shown on the right.}
\label{fig:sing_het_kl_zT}
\end{figure}

Results for $\kappa_l$ are shown in Fig.~\ref{fig:sing_het_kl_zT} (top
panel) for the same single Ge heterostructure SiNWs discussed
earlier. We find that, by introducing a single Ge heterostructure, the
lattice thermal conductivity can be reduced by a factor of five for the \d111
growth direction, as compared to the corresponding pristine SiNW,
giving a value of 0.1~nWK$^{-1}$. Reductions in the \110 and \211
direction are also significant (approximately a factor of four).  
We tentatively suggest that the longer unit cell in the \d111
direction may account for the greater reduction seen in that growth
direction.
It may also be seen that, in the \110 direction, $\kappa_l$ 
increases with diameter as more phonon modes become available.
We also observe this trend in larger diameter pristine SiNWs in the
\d111 and \211 directions (results not shown). 

\section{$zT$ for SiNWs with a single Ge heterostructure}
Fig.~\ref{fig:sing_het_kl_zT} (bottom panel) combines our results for
electronic and phononic transport coefficients for the single Ge
heterostructure systems discussed above and shows our calculated
values of \ztm\, as a function of heterostructure length. 
It can be seen that heterostructured SiNWs in the \d111 direction
display the greatest values of the figure of merit \ztm~$\simeq
1.4$. Such high values, however, are not found consistently across the
range of heterostructures studied. 
Such variations may limit the values of  $zT$ observed in realistic
SiNWs since experimental control over heterostructure length is,
currently at least, limited to length-scales comparable to, or greater
than, the differences in length that are investigated here~\cite{wen_2009}.
In the \110 and \211 directions, we find \ztm~$< 1$, mainly due to the
higher lattice thermal conductivities found in these systems.
We note that, across the range of systems studied, the ratio of
lattice and electronic thermal conductances, $\kappa_l/\kappa_e$, lies
between 3 and 10, therefore, $\kappa_l$ is the dominant contribution
to the denominator of $zT$. This emphasizes the importance of reducing
the lattice thermal conductivity for high $zT$ NWs.

\section{$zT$ for SiNWs with multiple Ge heterostructures}

\begin{figure}
\onefigure[height=2.4cm]{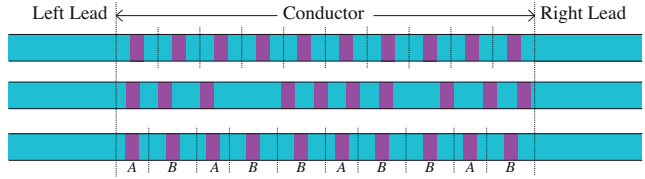}
\caption{Schematic illustration of MHNWs studied. Top: periodic
  conductor; middle: a typical random arrangement; bottom:  
Fibonacci chain pattern, with units $A$ and $B$ (see text). Si
and Ge sections are in blue and magenta, respectively.}
\label{fig:MHNWs}
\end{figure}

Next, we consider much longer SiNWs
with many Ge heterostructures 
along their length. Such multiple heterostructure nanowires (MHNWs)
are shown schematically in Fig.~\ref{fig:MHNWs}. These systems are too
large for brute-force PW-DFT calculations. Instead, we use
the Hamiltonian matrices of single heterostructure calculations as
``building-blocks'' for constructing model 
Hamiltonians of much larger (up to $\simeq8400$ atom) MHNWs, with
negligible loss of accuracy. Our
approach, which relies on exploiting the nearsightedness of electronic
structure that becomes manifest when Hamiltonian matrices are
represented in a basis of MLWFs, is described in detail in
Ref.~\cite{shelley_method2011}. 

Once the model Hamiltonian for the MHNW is constructed, 
the electronic transport properties 
under open boundary conditions 
are
calculated in exactly the same way as described above for SiNWs with single
Ge heterostructures. 
For the lattice thermal conductivity, an analogous ``building-block'' 
scheme is used in which the (short-ranged) dynamical matrices of the
single heterostructures that comprise the MHNWs are combined to
construct dynamical matrices for the MHNWs. Under the assumption that
phonons remain phase coherent across the length of the MHNW, this
dynamical matrix is then used to calculate the phonon transmission
function $\mathcal{T}(\omega)$ and, hence, the coherent lattice
thermal conductance $\kappa_l^\mathrm{coh}$ according to
Eq.~(\ref{eq:kappa_l}).

It is unclear whether the phase coherence length of phonons is comparable
to the lengths of the MHNW systems that we consider (up to
$\simeq$~116~nm). Therefore, we also calculate the lattice thermal
conductance in an ohmic regime, $\kappa_l^\mathrm{ohm}$, in which the
total resistance of a given MHNW is the sum of the thermal resistances
of each individual heterostructure that constitutes the
MHNW~\cite{markussen-prl07}. 
For a NW with $N$ heterostructures, each of which in
  isolation gives a transmission $\mathcal{T}_i(E)$, we compute the
  transmission function as~\cite{markussen_2009}
\begin{equation}
\mathcal{T}^{\mathrm{ohm}}(E) = \frac{\overline{\mathcal{T}}}{N -
  (1-N)\overline{\mathcal{T}}/\mathcal{T}_0}, 
\end{equation}
where $\overline{\mathcal{T}} = \sum_{i=1}^{N} \mathcal{T}_i / N$ is
the average transmission of the isolated heterostructures, and
$\mathcal{T}_0$ is the transmission for the pristine
NW. Having obtained $\mathcal{T}^{\mathrm{ohm}}(E)$,
$\kappa_{l}^{\mathrm{ohm}}$ is calculated from Eq.~(\ref{eq:kappa_l}).
$\kappa_l^\mathrm{coh}$ and $\kappa_l^\mathrm{ohm}$ are used to
estimate upper and lower bounds for $zT$.

We consider MHNWs in the \d111 and \211 growth directions, with
diameters of 1.14~nm and 1.06~nm, respectively, and Ge heterostructure 
lengths of 3.80~nm and 1.34~nm, respectively.
For each of these two growth directions, three qualitatively distinct
heterostructure distributions are considered (shown in
Fig.~\ref{fig:MHNWs}): (i) random, (ii) periodic, and (iii) Fibonacci
chain\footnote{A Fibonacci chain is an example of a 1D 
  quasicrystal~\cite{levine_1984}: it displays local translational
  symmetries, yet remains aperiodic \emph{in toto}. Exceptionally low
  $\kappa_l$ values have been reported experimentally for
  3D quasicrystals~\cite{pope_2004}, thus the
  introduction a Fibonacci chain distribution of heterostructures
  could be a systematic method to reduce $\kappa_l$. The Fibonacci
  chain MHNWs are designed such that the length ratio of structural
  units $A$ and $B$ that comprise them is as close as possible to the
  golden ratio $(1+\sqrt{5})/2$. These structural units each contain a
  Ge heterostructure between lengths of SiNW and the total chain is
  built with three iterations ($n=0,1,2$) of the sequence:
  $A_{n+1}=A_nB_n$, $B_{n+1}=A_n$, with $A_0=A$ and $B_0=B$.}, each
with ten heterostructures along the length of the NW. 
Structural details are given in Table~\ref{tab:LDC_comparison}. 
For MHNWs with a random distribution of heterostructures,
$\T(\epsilon)$ and $\T(\omega)$ are ensemble averaged
over 250 independent configurations.

\begin{table}
\small
\centering
\begin{tabular}{l|l|c|c}
\multicolumn{2}{c|}{} & \d111 & \211 \\
\hline 
\hline 
\multicolumn{2}{r|}{Single heterostructure length (nm)} 
& 3.80 & 1.34 \\ \hline
Periodic
&Total MHNW length	(nm) & 93.3		& 50.9		\\
&	Total number of atoms	& 	7208	& 3608		\\
 \cline{1-4}
Random
&Total MHNW length	(nm) & 93.3		& 49.6		\\
&	Total number of atoms	& 	7208	& 	3520		\\
\cline{1-4}
Fibonacci
&Total MHNW length (nm)	& 93.3		& 49.6		\\
&	Total number of atoms	& 	7208	& 	3520		\\
\cline{1-4}
\hline
\hline
\end{tabular}
\caption{Structural details of the MHNWs in the \d111 and \211 growth
  directions. The different MHNWs 
  (periodic, Fibonacci and random) are built by placing ten single
  heterostructures along the length of the wire.} 
\label{tab:LDC_comparison}
\end{table}

\begin{figure}
\onefigure[width=7.5cm,angle=270]{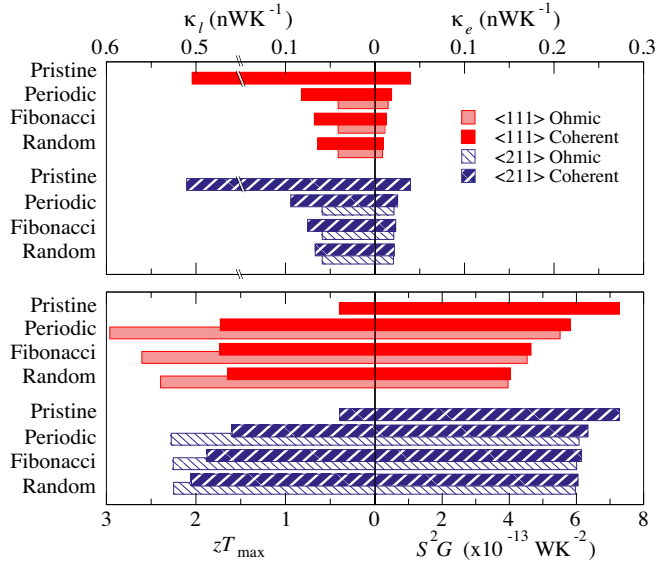}
\caption{Transport properties at 300~K of \d111
  (red, solid bars) and \211 (blue, striped bars) periodic, 
Fibonacci and random patterned MHNWs, and comparison
to the pristine cases. Results from fully coherent phononic transport
\d111 (\211) calculations a have solid shading (bold stripe), while
results in the ohmic regime have a lighter shading (fine stripe). Top
panel: the lattice (left) and electronic (right) contributions to the 
thermal conductance, respectively. Bottom panel: \ztm\ 
(left) and \s2g\ (right) at the value of $\mu$ that maximizes $zT$ for 
each system.} 
\label{fig:LDC_zTmax}
\end{figure}

Fig.~\ref{fig:LDC_zTmax} (top panel) shows the thermal conductances of
the MHNWs detailed in Table~\ref{tab:LDC_comparison} and compares them
to the pristine values. A prominent feature is the large reduction of
$\kappa_l$ due to heterostructuring, with \d111 MHNWs displaying
smaller values than \211. Using the coherent model (red solid
bars/blue bold striped bars), $\kappa_l$ is reduced by factors of
between five and eight when compared to the pristine results and
reduces as the disorder is increased (periodic to Fibonacci to random
patterning). 
The ohmic model (red shading/fine blue stripes), results in
reductions of $\kappa_l$ by factors of $\simeq$~12 and 8.5 in \d111
and \211 MHNWs, respectively. We note that
$\kappa_l^\mathrm{coh}/\kappa_e$ and $\kappa_l^\mathrm{ohm}/\kappa_e$
are found to be between 2.5 and 7 --- only marginally smaller than
those values we obtained for single heterostructure NWs.

\ztm\ and \s2g\ are shown in Fig.~\ref{fig:LDC_zTmax} (bottom
panel).\footnote{The electronic properties are always calculated
  within a fully coherent model but, depending on whether
  $\kappa_{l}^{\mathrm{ohm}}$ or $\kappa_{l}^{\mathrm{coh}}$ is used,
  the value of $\mu$ at which \ztm\ occurs changes slightly and,
  hence, the value of \s2g, which is dependent on $\mu$.}
It is striking that in no case does \s2g\ increase due to
heterostructuring,
and the \211 direction performs best over the range of MHNWs
considered, showing only small decreases (with respect to pristine) as
the disorder increases from periodic to random.
Together with the pronounced effect that increased disorder has on 
$\kappa_l^\mathrm{coh}$, we see that, in the coherent regime, random
patterning results in values of the figure of merit as high as
$zT=2$. Conversely, \d111 MHNWs display significant 
reductions in \s2g\ as the disorder increases, which tend to
counteract similar decreases in $\kappa_l^\mathrm{coh}$, leaving \ztm\
approximately constant at $\simeq$~1.7. 
In both \d111 and \211 MHNWs the calculated \ztm\ increases if the
thermal transport is assumed to be ohmic, since
$\kappa_l^\mathrm{ohm}<\kappa_l^\mathrm{coh}$.
In this regime, $\kappa_l^\mathrm{ohm}$ is invariant with respect to
the distribution of heterostructures, therefore, \ztm\ in this regime
will follow the behaviour seen in \s2g\, with a value of $zT=2.3$ for 
the \211 direction (almost independent of the distribution of
heterostructures), and up to $zT=3$ in the \d111 direction, with a
periodic arrangement of heterostructures.

\section{The effect of variability of heterostructure length}
The MHNWs that we have discussed thus far consist of multiple
instances of identical heterostructures. 
Experimental synthesis techniques do not have this level of atomic
precision, therefore, we have investigated the effect of introducing
some variability of the lengths of the heterostructures that comprise
the MHNW.
In particular, we compare periodic patterned MHNWs with identical
heterostructures with `near-periodic' MHNWs that are comprised of
heterostructures whose lengths are Gaussian distributed about a mean
length that is given by the heterostructure length used in the `true'
periodic case,
with a standard deviation $\sigma$ that corresponds to approximately
1/3 (1/2) a unit cell in the \d111 (\211) direction. 
For the near-periodic MHNWs, the transmission functions are ensemble
averaged over 250 independent configurations of the disorder to model
a `typical' MHNW of this type.

\begin{figure}
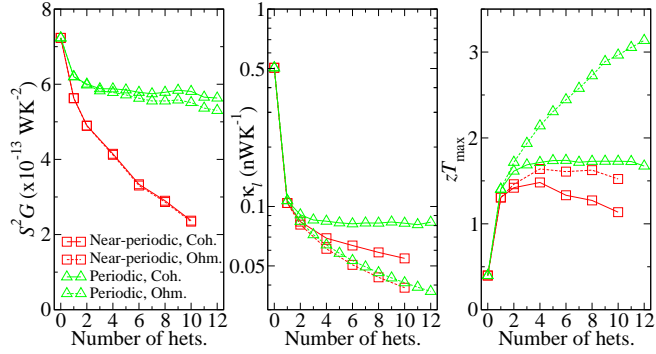

\centering
\onefigure[width=8.5cm]{var_het_v_per2.eps}
\caption
{Comparison of \s2g\ (left panel), $\kappa_l$ (middle panel) and  \ztm\ 
(right panel) at 300~K in periodic (green triangles) and near-periodic (red squares)
\d111 MHNWs as a function of the number of Ge heterostructures (hets.).
Results using the coherent (coh.) and ohmic (ohm.) phonon transport models are shown
with solid and dashed lines respectively.} 
\label{fig:var_het_v_per}
\end{figure}

We consider MHNWs with up to twelve heterostructures along their length. 
Fig.~\ref{fig:var_het_v_per} shows a comparison between near-periodic
(red squares) and periodic (green triangles) MHNWs in the \d111 growth
direction.
The resulting values for \ztm\ are shown in the right panel. In the
ohmic phonon transport regime (dashed lines), the near-periodic system
displays a dramatic reduction in \ztm\ as compared to the periodic
case, which arises from the sharp reduction that is found in \s2g\
(left panel), combined with the fact that $\kappa_l$ does not decrease
very much (middle panel).  
When considering the coherent regime (solid lines), the reduction in
\s2g\ for the near-periodic MHNW is also large (left panel), as
compared to the periodic MHNW, but associated decreases in $\kappa_l$
are also observed (middle panel) so that the resultant drop in \ztm\
(right panel), as compared to the periodic case, is much less
pronounced. For the near-periodic MHNWs, both phonon transport regimes
display a maximum in \ztm\ with respect to heterostructure length
after the introduction of approximately four heterostructures. 

We note that the significant decreases in \s2g\ due to the
variability in the heterostructure length is
consistent with our earlier conclusion that increased disorder tends
to reduce the power factor, as was seen when comparing periodic,
Fibonacci and random distributions. 
We also note that in the ohmic phonon transport regime, there is
little difference in $\kappa_l$ between periodic and near-periodic
MHNWs, which follows the earlier observation that there is almost no
dependence of $\kappa_l$ on heterostructure length in single
heterostructure NWs (Fig.~\ref{fig:sing_het_kl_zT}, top panel).

Finally, comparing near-periodic MHNWs in the \d111 and \211
  growth directions, we find that $S^2G^{\langle 111\rangle}<S^2G^{\langle
  211\rangle}$, which may have been expected from the stronger 
dependence of \s2g\m\ on heterostructure length in \d111
SiNWs with a single heterostructure (Fig.~\ref{fig:sing_het_S2G}). We
also find that  $\kappa_l^{\langle 111\rangle}<\kappa_l^{\langle
  211\rangle}$, which also could have been predicted from the trends
observed for single heterostructure SiNWs
(Fig.~\ref{fig:sing_het_kl_zT}). However, the 
delicate balance between \s2g\ and $\kappa_l$ make it difficult to
use calculations on SiNWs with a single heterostructure to predict
trends in \ztm\ for our near-periodic MHNWs, highlighting the need for
accurate first-principles approaches. Among the near-periodic
MHNWs studied, the \d111 direction with four heterostructures
performed best, with $zT\simeq 1.5-1.6$.

In conclusion, we have performed first-principles calculations on
thin, $p$-type \110, \d111 and \211 SiNWs with Ge 
heterostructures. 
In all cases studied, a decrease of thermoelectric power factor \s2g\
is observed when a heterostructure is introduced, and any increase
in the figure of merit $zT$ is due to a corresponding reduction in
$\kappa_l$. 
We have built model Hamiltonians for MHNWs with
over 8400 atoms while retaining first-principles accuracy. A similar
method was applied to the dynamical matrices of MHNWs to obtain the
thermal conductance $\kappa_l$ for such structures.
In such MHNWs we again find that \s2g\ is always reduced and that
increases in $zT$ are driven predominantly by significant decreases in
$\kappa_l$. We find values as high as $zT=3$ in \d111 MHNWs with
periodic arrangements of Ge heterostructures. The intricate
balance between \s2g\ and $\kappa_l$, however, makes $zT$ strongly
dependent on the details of the system at the atomic level: in
structures that model the kind of disorder that may be present in
realistic MHNWs, more modest values of $zT=1.6$ are found, which is 
still a factor of four greater than the pristine SiNW case. Our
calculations suggest axial heterostructuring to be a promising route
to high-$zT$ nanowire thermoelectrics.

\acknowledgments
We are grateful to the High Performance Computing Facility at Imperial
College London, and to the EPSRC and E.ON's International Research
Initiative. We thank N.~Poilvert, N.~Marzari and Y.-S.~Lee for
discussions.

\end{document}